\documentclass[epj,final]{svjour}
\usepackage{amssymb}
\usepackage{epsfig}
\usepackage{amsmath,amsfonts,latexsym}
\DeclareMathOperator{\Tr}{Tr}
\begin{document}
\title{Prospects of Open Charm Production\\ at GSI-FAIR and J-PARC}
\author{J.\ Riedl\inst{1}, A.\ Sch\"{a}fer\inst{1} \and
M.\ Stratmann\inst{2}
}                     
\institute{\inst{1}
Institut f{\"u}r Theoretische Physik, Universit{\"a}t Regensburg,
D-93040 Regensburg, Germany\\
\inst{2} Radiation Laboratory, RIKEN, 2-1 Hirosawa, Wako, Saitama 351-0198, Japan}
%
\date{}
%
\abstract{We present a detailed phenomenological study of the prospects of
open charm physics at the future $\bar{p}p$ and $pp$ facilities  
GSI-FAIR and J-PARC, respectively.
In particular, we concentrate on differential cross sections and the charge and
longitudinal double-spin asymmetries at next-to-leading order accuracy.
Theoretical uncertainties for the proposed observables are estimated by
varying the charm quark mass and the renormalization and factorization scales.
\PACS{
      {12.38.Bx}{}   \and
      {13.88.+e}{}   
     } 
} 

\maketitle
\section{Motivation and Introduction}
In recent years, the study of heavy flavors at colliders has become a versatile tool to probe different
aspects of Quantum Chromodynamics (QCD), ranging from heavy flavor parton densities and the hadronization
of heavy quarks into heavy mesons or baryons to the dynamics of QCD hard-scattering \cite{ref:hqref}.
Technological advances, like the introduction of micro-vertex detectors,
allow for much better tagging of the produced heavy quarks and hence more precise measurements.
An equally significant improvement of the theoretical foundations for heavy flavor production has been achieved,
in particular, novel proposals to overcome limitations of fixed-order calculations and to model the hadronization
of heavy quarks more reliably \cite{ref:hqref}.
As a consequence, a discrepancy between data and theory for bottom production,
suggested by various experiments, has been reduced
to the point that it no longer appears significant \cite{ref:cacciari-bottom}.

A solid understanding of the theoretical framework for heavy quark production
and its phenomenological inputs is of utmost importance for several key measurements both
ongoing or taking place in the very near future \cite{ref:hqref}.
At the LHC, heavy flavor production by genuine QCD processes is an important
background to searches for Higgs bosons within the Standard Model and beyond,
e.g., to the decay $H\to b\bar{b}$, as well as for new physics.
At the Relativistic Heavy Ion Collider (RHIC) one wants to establish the existence and to investigate
the properties of a new state of matter, the quark-gluon plasma (QGP). Several signatures related
to heavy flavors have been proposed, in particular, how their production and decays are modified by
the presence of a QGP \cite{ref:rhicqgp}.
Here, heavy flavor production in proton-proton collisions will act as a
benchmark for possible modification in nucleus-nucleus 
collisions at the same energy \cite{ref:rhicqgp,ref:cacciari-rhic}.
In addition, RHIC is also capable to collide longitudinally polarized protons at center-of-mass system
(c.m.s.) energies of up to $\sqrt{S}=500\,\mathrm{GeV}$. Besides its intrinsic interest as an important
test of the dynamics of spin-dependent hard-scattering in QCD, these measurements are
likely to further our understanding of how gluons contribute to the proton spin,
a major goal of hadron physics \cite{ref:spinplan}.

In terms of perturbative QCD (pQCD), relevant fixed-order calculations have been in place for quite
some time now, and next-to-leading order (NLO) accuracy is the state-of-the-art throughout. NLO
results, which keep the full dependence on the heavy quark mass $m_Q$,
for unpolarized hadron-hadron (``hadroproduction''), photon-hadron (``photoproduction''), and electron-hadron
(``elec\-tro\-pro\-duc\-tion'') collisions can be 
found in \cite{ref:nlohadro}, \cite{ref:nlophoto}, and \cite{ref:nloelectro},
respectively. Calculations for longitudinally polarized hadro- \cite{ref:polnlohadro}
and photoproduction \cite{ref:polnlophoto} are more recent achievements.
In each case one exploits the fact that $m_Q$ acts as an effective infrared cut-off
for collinear singularities, which allows to compute total heavy quark yields as a perturbative series in
the strong coupling $\alpha_s$ evaluated at a hard scale of ${\cal{O}}(m_Q)$.
For inclusive transverse-momentum $p_T$ distributions, complications arise
once $p_T/m_Q\gg 1$, and large logarithms in this ratio have to be
resummed to all orders in $\alpha_s$ to improve the convergence of the perturbative series \cite{ref:cacciari-resum}.

In this paper, we will focus on a detailed phenomenological study of the prospects of open charm production in
$\bar{p}p$ and $pp$ collisions at the future GSI-FAIR \cite{ref:gsi-fair} and J-PARC \cite{ref:jparc}
facilities, respectively.
For GSI-FAIR we consider the so-called ``collider option'' as proposed by the PAX collaboration \cite{ref:gsi-pax},
using a $15\,\mathrm{GeV}$ anti-proton and a $3.5\,\mathrm{GeV}$ proton beam,
with the plan of having both beams also longitudinally or transversely polarized.
The main goal of the PAX experiment would be a determination of the 
so far unknown ``transversity'' parton densities
through the transversely polarized Drell-Yan process. This measurement would gain substantially from
a polarized anti-proton beam due to the dominance of the lowest order (LO) quark-antiquark annihilation channel.
The J-PARC facility in Japan is currently being completed. Here we consider the collision of a
$50\,\mathrm{GeV}$ proton beam with a fixed, solid-state target. The possibility of
having both beam and target polarized is a conceivable option for future upgrades currently being
scrutinized. At GSI-FAIR (J-PARC) the available
$\bar{p}p$ $(pp)$ c.m.s.\ energy will be $\sqrt{S}\simeq 14.5\,(10)\,\mathrm{GeV}$.
For both experiments the details of the detector and the acceptance are not yet finalized. We will make
some reasonable assumptions as stated below in Sec.~\ref{sec:inputs}.
We will demonstrate that measurements of open charm at these facilities have the potential to
further our understanding of the underlying QCD dynamics at moderate c.m.s.\ energies so far little explored.

First, one has to determine, of course, to what extent perturbative QCD is applicable at c.m.s.\ energies 
of about 10 to 15~GeV. NLO calculations for single-inclusive $p_T$-spectra of pions or photons 
are known to seriously undershoot data even at somewhat higher c.m.s.\ energies \cite{ref:werner-resum}.
The mass of the charm quark already sets a hard scale of ${\cal{O}}(1\,\mathrm{GeV})$, which may
facilitate the use of perturbative QCD even for small or vanishing transverse momenta $p_T$ of
the observed charm quark.
Also, since $p_T\approx {\cal{O}}(m_Q)$, we do not have to worry about
potentially large logarithms $\ln p_T/m_Q$ present at large c.m.s.\ energies \cite{ref:cacciari-resum},
however, partonic threshold effects may become important and perhaps need to be resummed to all orders.
In case of $p_T$-differential pion spectra, it was shown that threshold resummations can lead to a much
improved agreement between theory and experiment at low c.m.s.\ energies \cite{ref:werner-resum}.
To study the possible relevance of resummations at GSI-FAIR and J-PARC, 
we provide total and differential charm yields 
for both $\bar{p}p$ and $pp$ collisions at NLO accuracy of QCD,
including detailed discussions of theoretical uncertainties due to variations of the charm quark mass
or the renormalization and factorization scales. We note that the computation of threshold resummations
for charm production at GSI-FAIR and J-PARC is far beyond the scope of this paper, which aims at
a first exploratory study of charm physics at comparatively low c.m.s.\ energies.

Other interesting, though experimentally more challenging observables related to open
charm production can be considered at GSI-FAIR and J-PARC.
One is the so called ``charge asymmetry'' which describes the difference of cross sections 
for producing a heavy quark $Q$ or a heavy anti-quark $\bar{Q}$ at a certain point in phase-space:
\begin{equation} \label{eq:chargeasy}
A_C\equiv\frac{d\sigma^Q -d\sigma^{\bar{Q}}} {d\sigma^Q + d\sigma^{\bar{Q}}}\;\;\;.
\end{equation}
This asymmetry probes a subset of NLO radiative corrections and vanishes at the LO
approximation. This feature makes it an important test of QCD hard scattering dynamics.
The Abelian (QED) part of $A_C$ is also known as forward-backward asymmetry \cite{ref:qed} 
and is caused by the interference of states with different $C$-parity.
It should be mentioned that routinely used event generators based on LO matrix elements
cannot predict this interesting effect.
In the unpolarized case the charge asymmetry was first mentioned in \cite{ref:charge-old,ref:nlohadro}
and later studied qualitatively in \cite{ref:kuhn,ref:topasy}, mainly for top production at the TeVatron and the LHC.
A first measurement of $A_C$ for top production has been recently reported by 
the CDF collaboration \cite{ref:cdfasy}.
We estimate the size of this effect for GSI-FAIR and J-PARC and, for the first time, compute
the corresponding charge asymmetry also for polarized hadroproduction.
We will give more details in Sec.~\ref{sec:chargeasym}.

Provided that longitudinally polarized beams and targets
will be available at GSI-FAIR and/or J-PARC, studies of the double-spin asymmetry
\begin{equation}
\label{eq:all}
A_{LL}\equiv\frac{d\Delta\sigma}{d\sigma}.
\end{equation}
could provide unique insight in the distributions describing the polarization of quarks and gluons 
in the nucleon at medium-to-large momentum fractions $x$. In (\ref{eq:all}), $d\Delta\sigma$ denotes the
spin-dependent cross section, which can be obtained by taking the difference of measurements
with the (anti-)proton spins aligned and anti-aligned. The
unpolarized cross section $d\sigma$ is determined by the sum of both measurements.
We will explore the sensitivity of $A_{LL}$ to different assumptions about polarized parton densities,
in particular, the elusive gluon polarization.
As for the unpolarized cross sections, we discuss theoretical uncertainties related to
variations of the charm quark mass and the  renormalization and factorization scales.

The paper is organized as follows: in Sec.~2 we briefly review the technical framework for
hadroproduction of heavy flavors, set our notation, and introduce in some detail the charge asymmetry.
Section 3 is devoted to numerical studies. First we define the phenomenological inputs and
experimental cuts we assume for GSI-FAIR and J-PARC. Next, we give results for total and
differential unpolarized charm cross sections and discuss their theoretical uncertainties.
These results serve as a future benchmark to study the applicability of pQCD for charm
production at $\sqrt{S}\simeq 10\,\mathrm{GeV}$. Then we present expectations for the
polarized cross sections and charge and double spin asymmetries.
We briefly summarize the main results in Sec.~4. 

\section{Technical Framework}
The pQCD framework for single-inclusive heavy flavor production at NLO accuracy
in both unpolarized and longitudinally polarized hadron-hadron collisions
has been discussed in detail in Refs.~\cite{ref:nlohadro} and \cite{ref:polnlohadro}, respectively.
We can restrict ourselves to a brief summary of the aspects 
with particular relevance to our analysis.

\subsection{The single-inclusive cross section}
We are interested in the hadroproduction of a heavy quark $Q$
[anti-quark $\bar{Q}$] with mass $m_Q$ and four-momentum $P_Q$ [$P_{\bar{Q}}$]:
\begin{equation}
\label{eq:hhqx}
 H_1(P_1) H_2(P_2)\to Q(P_Q) [\bar{Q}(P_{\bar{Q}})]+X.
\end{equation}
$X$ includes all other final-state particles such that (\ref{eq:hhqx})
is inclusive with respect to the detected heavy quark.
$P_{1,2}$ denote the momenta of the colliding hadrons $H_{1,2}$
and $S=(P_1+P_2)^2$ the c.m.s.\ energy squared.

More specifically, $P_Q$ can be parameterized in terms of
the transverse momentum $p_T$, the rapidity $y$, and the
azimuthal angle $\phi$ of the observed heavy quark $Q$,
\begin{equation}
\label{eq:fourmom}
P_Q=(m_T \cosh y, p_T \sin\phi, p_T \cos\phi, m_T \sinh y),
\end{equation}
where the transverse mass is defined as $m_T=\sqrt{p_T^2+m_Q^2}$
and $y=\frac{1}{2} \ln[(E+P_z)/(E-P_z)]$.
Applying the factorization theorem,
the unpolarized differential cross section for (\ref{eq:hhqx})
can be written schematically as
\begin{eqnarray}
\label{eq:diffxsec}
\nonumber
\frac{d^2\sigma^Q}{d m_T^2 dy} &=& \sum_{ab} f_a^{H_1}(x_1,\mu_f)
\otimes f_b^{H_2}(x_2,\mu_f) \\
&&\otimes\, \frac{d^2\hat{\sigma}_{ab\to QX}(x_1,x_2,P_1,P_2,P_Q,\mu_f,\mu_r)}{d m_T^2 dy}
\end{eqnarray}
with the symbol $\otimes$ indicating a convolution.
$x_1$ and $x_2$ are the fractions of $P_1$ and $P_2$ taken by the
partons $f_a^{H_1}$ and $f_b^{H_2}$ with flavor $a$ and $b$, respectively.
The sum in Eq.~(\ref{eq:diffxsec}) is to be performed over all
contributing partonic channels $ab\to Q X$ with $d\hat{\sigma}_{ab\to Q X}$ the
associated partonic cross sections.

The factorization of the cross section (\ref{eq:diffxsec}) into non-perturbative
parton densities $f_{a,b}$ and short-distance cross sections
requires the introduction of factorization and re\-nor\-malization sca\-les $\mu_f$ and $\mu_r$, respectively.
These scales are essentially arbitrary and usually chosen to be of the
order of a hard momentum transfer characterizing the process under consideration.
Any residual dependence of the right-hand-side (r.h.s.) of Eq.~(\ref{eq:diffxsec})
on the actual choice for $\mu_{f,r}$ represents an important part of the
uncertainties in the theoretical description of (\ref{eq:hhqx}).

The basic framework outlined above carries over to the case of polarized
hadron-hadron collisions as well. To obtain the spin-dependent
cross section $d^2\Delta\sigma/dm_T^2 dy$ entering the
experimentally relevant spin asymmetry $A_{LL}$ defined in Eq.~(\ref{eq:all}),
the parton densities and hard scattering cross sections
on the r.h.s.\ of Eq.~(\ref{eq:diffxsec})
have to be replaced by their polarized counterparts
$\Delta f_{a,b}$ and $d\Delta\hat{\sigma}_{ab\to Q X}$, respectively.

Knowledge of higher-order corrections in the perturbative expansion of the
partonic cross sections $d\hat{\sigma}_{ab\to Q X}$ and
$d\Delta\hat{\sigma}_{ab\to Q X}$ is generally indispensable. On the one hand, in
hadronic scattering they are often sizable, and, on the other hand, they
are expected to reduce the artificial dependence on the choice of $\mu_{f,r}$.
Also, higher-order corrections affect unpolarized and polarized cross sections
differently and hence do not cancel in the ratio $A_{LL}$.

At ${\cal O}(\alpha_s^2)$, the LO approximation, the hadroproduction of
heavy quarks proceeds through only two partonic channels
\begin{equation}
\label{eq:loproc}
gg\to QX\,\, \text{and}\,\, q\bar{q}\to Q X
\end{equation}
where $X=\bar{Q}$. The ${\cal O}(\alpha_s)$ radiative
corrections to (\ref{eq:loproc}) comprise additional
real gluon emission, $X=\bar{Q}g$,  as well as
one-loop (virtual) contributions. In addition,
a new type of subprocess, gluon-[anti]quark-scattering,
$gq [\bar{q}]\to Q X$ has to be considered at NLO.
A detailed account of the calculation of the
relevant matrix elements, the required loop and phase-space
integrations, and the cancellation of singularities
is given in Refs.~\cite{ref:nlohadro} and \cite{ref:polnlohadro}.

Since $m_Q$ acts as an effective cut-off for collinear singularities,
also total heavy quark yields are amenable to pQCD.
They are obtained by integrating (\ref{eq:diffxsec}) over
the entire phase-space using
\begin{eqnarray}
\label{eq:totalxsec}
\nonumber
&&\int_{m_Q^2}^{S/4} dm_T^2 \int_{-\cosh^{-1} \sqrt{S}/(2m_T)}^{\cosh^{-1} \sqrt{S}/(2m_T)} dy = \\
&&\int_{-\frac{1}{2} \ln \frac{1+\beta}{1-\beta}}^{\frac{1}{2} \ln \frac{1+\beta}{1-\beta}} dy
\int_{m_Q^2}^{S/(4\cosh^2 y)} dm_T^2,
\end{eqnarray}
where $\beta\equiv\sqrt{1-4m_Q^2/S}$. Alternatively, one can first derive the so-called
``scaling functions'', $[\Delta]\hat{\sigma}_{ab}(s,m_Q^2)$,
as a function of the partonic c.m.s.\ energy $s$ by integrating the
partonic cross sections $d[\Delta]\hat{\sigma}_{ab\to Q X}$ \cite{ref:nlohadro,ref:polnlohadro},
which then in turn have to be combined with the appropriate combination of parton densities.
We note that the experimental determination of the total cross section always
involves some extrapolation beyond the accessible ranges in $p_T$ and $y$ and
is therefore less reliable and useful than differential rates in testing pQCD predictions.
In our numerical studies for J-PARC and GSI-FAIR we will therefore mainly focus on
differential cross sections, which are also relevant for the charge asymmetry.

\subsection{The charge asymmetry \label{sec:chargeasym}}
%
At the LO approximation the processes (\ref{eq:loproc}) relevant for heavy flavor production
do not discriminate between a produced heavy quark $Q$ and a heavy anti-quark $\bar{Q}$.
Hence, at any given point $(p_T,\,y)$ in phase-space the yields (\ref{eq:diffxsec}) for
$Q$ and $\bar{Q}$ are identical.

Radiative corrections change this picture and give rise to the charge asymmetry $A_C$,
defined in Eq.~(\ref{eq:chargeasy}). Any measurement of $A_C$ will
directly probe and perhaps improve our understanding of QCD dynamics beyond the LO.
So far this higher-order effect has received relatively
little attention \cite{ref:charge-old,ref:kuhn,ref:topasy} but was recently measured
for the first time in case of top production at CDF \cite{ref:cdfasy}, the asymmetry 
exceeding the few percent predicted in \cite{ref:kuhn} with large experimental
uncertainties though.
We will explore the prospects of accessing $A_C$ in $pp$ and $\bar{p}p$ collisions
at J-PARC and GSI-FAIR, respectively. Due to the relatively low c.m.s.\ energies
available, we have to limit ourselves to charm quark production.
Given that longitudinally polarized beams and targets are a viable option, 
we also explore the polarized charge asymmetry,
which we define as in Eq.~(\ref{eq:chargeasy}) with the cross sections $d\sigma^{Q [\bar{Q}]}$
replaced by their polarized counterparts $d\Delta\sigma^{Q [\bar{Q}]}$.

\begin{figure}[t!]
\begin{center}
\epsfig{figure=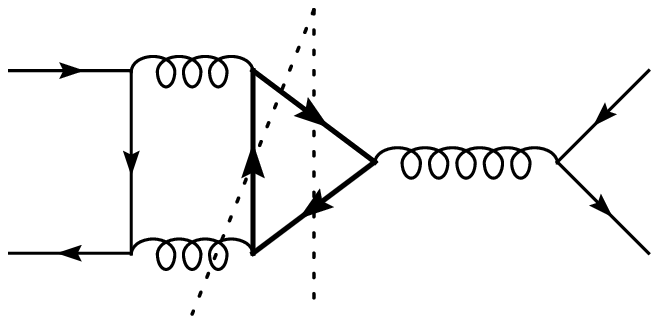,width=0.3\textwidth}
\end{center}
\caption{Sample $q\bar{q}$ cut diagram contributing to $A_C$.
\label{fig:qq-feyn}}
\begin{center}
\epsfig{figure=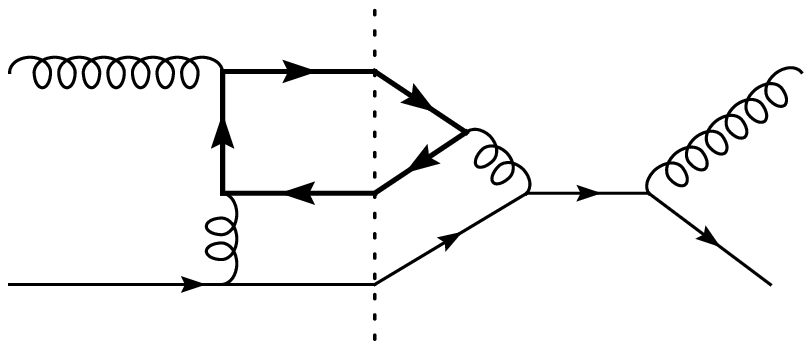,width=0.3\textwidth}
\end{center}
\caption{Sample $gq$ cut diagram contributing to $A_C$.
\label{fig:qg-feyn}}
\end{figure}

At NLO the charge asymmetry receives only a very limited number of contributions.
Instead of making use of the results available in the literature \cite{ref:nlohadro,ref:polnlohadro},
we chose to re-calculate the numerator of $A_C$ from scratch in both the unpolarized and polarized 
case\footnote{The lengthy expressions are available upon request from the authors.}.
The origin of $A_C$ resides in the Abelian (QED) part of the higher order processes
and relates to the interference between amplitudes which are relatively odd under
the exchange of $Q$ and $\bar{Q}$. The gluon-gluon fusion process in (\ref{eq:loproc})
is evidently charge symmetric also beyond the LO and does not contribute to $A_C$.
Also non-Abelian amplitudes involving the triple-gluon vertex depend only on
$Q+\bar{Q}$ and lead to contributions which are symmetric under exchanging $Q$ and $\bar{Q}$.
Upon close examination of the remaining amplitudes for the $q\bar{q}$ and $gq$ [$g\bar{q}$] initiated
subprocesses at NLO, one finds that only such ``cut diagrams'' contribute which have three
vertices on both the heavy and light quark line.
In case of the virtual loop corrections to the LO $q\bar{q}$ process, only the interference of the
box diagrams with the Born amplitude contributes.
Examples of cut diagrams relevant for the computation of $A_C$ and $\Delta A_C$
are depicted in Figs.~\ref{fig:qq-feyn} and \ref{fig:qg-feyn}.

This observation can be readily understood. To this end, let us write the partonic
subprocess cross section $d\hat{\sigma}_{ab\to QX}$ as the sum of
the interference contributions of all contributing amplitudes
labeled by the indices $i,j$
\begin{equation}
\label{eq:ac1}
d\hat{\sigma}_{ab\to QX} = K_{ab} \sum_{ij} {\cal{M}}_i {\cal{M}}_j^*\Big|_{ab\to QX}.
\end{equation}
For simplicity, pre-factors
such as the spin and color averages are mapped into $K_{ab}$. Phase-space integration
for all unobserved partons is implicitly understood in (\ref{eq:ac1}).
Expressions similar to (\ref{eq:ac1}) hold for
polarized partonic cross sections $d\Delta\hat{\sigma}_{ab\to QX}$
as well as for $ab\to \bar{Q}X$.
For the numerator of $A_C$ we have to examine the difference
\begin{equation}
\label{eq:ac2}
\Delta_{ij} = {\cal{M}}_i{\cal{M}}_j^*\Big|_{ab\to QX} -
{\cal{M}}_i{\cal{M}}_j^*\Big|_{ab\to \bar{Q}X}
\end{equation}
for all $i,j$.
By interchanging $Q$ and $\bar{Q}$, the Dirac structure relevant for  ${\cal{M}}_i{\cal{M}}_j^*$
changes sign for an odd number of propagators in the heavy quark trace and otherwise
remains the same. The color structure is invariant, except for
$\Tr[T_k T_l T_m]=(d_{klm}+if_{klm})/4$ which contains a symmetric
and an anti-symmetric piece, $d_{klm}=d_{mlk}$ and $f_{klm}=-f_{mlk}$, respectively.

Combining everything, only those cut diagrams with three vertices on both
the heavy and the light quark line, cf.\ Figs.~\ref{fig:qq-feyn} and \ref{fig:qg-feyn},
contribute to $A_C$, and one finds \cite{ref:kuhn}
\begin{equation}
\label{eq:ac3}
\Delta_{ij} = \frac{1}{8}(d_{klm})^2 \widetilde{{\cal{M}}_i{\cal{M}}_j^*}\Big|_{ab\to QX},
\end{equation}
with $(d_{klm})^2=40/3$. $\widetilde{{\cal{M}}_i{\cal{M}}_j^*}$ denotes the interference
of the two amplitudes with the QCD color structure taken aside, and is the same
(up to pre-factors) as for corresponding QED processes \cite{ref:qed}, e.g, $e^+ e^-\to \mu^+ \mu^- \gamma$.
Let us mention that there is a similar effect in the QCD scale evolution of parton densities
at next-to-next-to-leading order (NNLO) also proportional to $(d_{klm})^2$ and leading
to a strange quark asymmetry $s(x)\ne \bar{s}(x)$ \cite{ref:strangeasy}.

All contributions to the numerator of $A_C$ at ${\cal{O}}(\alpha_s^3)$ are free of ultraviolet as
well as collinear singularities as a consequence of the symmetry of the LO processes
(\ref{eq:loproc}) under exchanging $Q$ and $\bar{Q}$.
Infrared (IR) singularities appear in both real gluon emission and virtual loop corrections to
the LO $q\bar{q}$ process and cancel in the sum.
Effectively this implies that the NLO matrix elements only contain the charge asymmetry
at LO approximation.
As in \cite{ref:nlohadro,ref:polnlohadro} we use dimensional regularization to deal with
IR poles in intermediate steps of the calculation. In \cite{ref:kuhn} a small gluon energy 
$E_{cut}^g$ was used to cut off IR singularities. 
While the charge asymmetry also appears in the limit $m_Q\to 0$, it vanishes for the
total heavy quark cross section as a consequence of charge conjugation invariance.

In Sec.~\ref{sec:ac} we give some quantitative predictions for $A_C$ and $\Delta A_C$ at J-PARC and GSI-FAIR.
Due to the low c.m.s.\ energies, gluons are much less abundant than at high-energy colliders.
Since gluon-gluon fusion only contributes to the denominator of $A_C$, the prospects for
studying $A_C$ can be in fact more favorable at GSI-FAIR or J-PARC than at colliders like the TeVatron or the LHC.
Here, gluon-gluon fusion is by far the dominant mechanism for heavy quark production,
and one has to find suitable corners of phase-space to make $A_C$ experimentally accessible.
One important possibility is top quark production \cite{ref:kuhn,ref:topasy,ref:cdfasy},
which receives important contributions
from $q\bar{q}$ annihilation also at large $S$ thanks to the sizable $m_Q$.

\section{Phenomenological Applications}
%
\begin{figure}[th!]
\vspace*{-0.5cm}
\begin{center}
\epsfig{figure=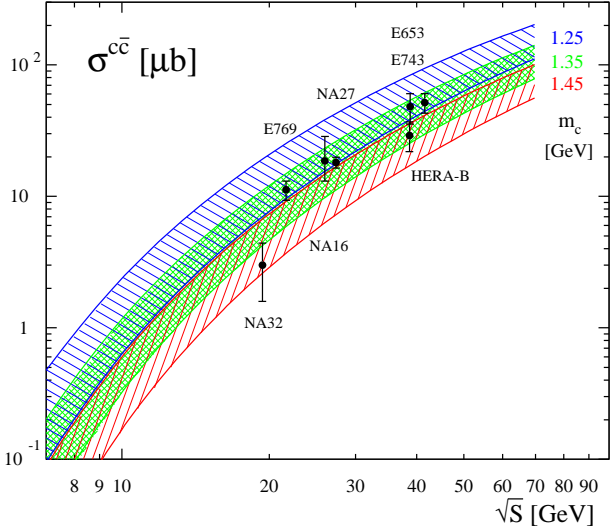,width=0.52\textwidth}
\end{center}
\caption{\label{fig:figure3} Experimental results \cite{ref:charm-review} for the total charm
production cross section at fixed-target energies compared to NLO pQCD calculations
for three different values of the charm quark mass $m_c$. In each case, the shaded band indicates the
theoretical uncertainties from varying $\mu=\mu_r=\mu_f$ in the range $m_c\le\mu\le 2m_c$.}
\end{figure}
Before turning to the prospects of heavy flavor physics at GSI-FAIR and J-PARC, we quickly review the
information on charm production in $pp$ collisions at c.m.s.\ energies below
$\sqrt{S}=50\,\mathrm{GeV}$ gathered so far.
Figure~\ref{fig:figure3} shows the available data \cite{ref:charm-review} compared to our
calculations at NLO accuracy using different values of the charm quark mass and scales
$\mu=\mu_f=\mu_r$ in Eq.~(\ref{eq:diffxsec}). As can be seen, uncertainties from small variations
of $m_c$ are as important as scale ambiguities, and both combined can lead to almost an
order of magnitude change in the total charm yield at $\sqrt{S}\simeq 10\div 20\,\mathrm{GeV}$.
From Fig.~\ref{fig:figure3} one can also infer that the theoretical uncertainties
become somewhat less pronounced with increasing c.m.s.\ energy.

Although most experimental results can be described with the same choice of $m_c$ and scale $\mu$,
there is a clear need for further precise measurements, in particular closer to threshold, below
the result of NA32, which does not line up so well with other experiments. GSI-FAIR and J-PARC,
to which we shall turn now, can explore this energy range in the future. 

\subsection{Phenomenological inputs and experimental cuts}\label{sec:inputs}
The part of the future GSI-FAIR accelerator complex ame\-na\-ble to pQCD studies is an
asymmetric proton-antiproton collider option proposed by the PAX collaboration \cite{ref:gsi-pax}
with maximum beam energies for protons and antiprotons of $3.5\,\mathrm{GeV}$ and $15\,\mathrm{GeV}$,
respectively, resulting in a c.m.s.\ energy of about $\sqrt{S}=14.5\,\mathrm{GeV}$.
Studies of methods to polarize both beams either longitudinally or transversely are currently
pursued \cite{ref:gsi-pax}. The PAX detector will have nearly full azimuthal acceptance and a polar angle coverage
from 5 to 130 degrees is envisioned \cite{ref:gsi-pax,ref:cont-priv}.

The proton accelerator at J-PARC, which will reach up to $50\,\mathrm{GeV}$
beam energy, is currently under construction,
and the hadron physics programme will commence in the near future at 
a c.m.s.\ energy of about $\sqrt{S}=10\,\mathrm{GeV}$ with both beam and target being unpolarized.
We also consider longitudinally polarized collisions, which are a conceivable upgrade of the 
J-PARC facility in the future \cite{ref:jparc}.
Since not very much is known about the experimental set-up at this stage, we assume a forward spectrometer geometry
with a $200\,\mathrm{mrad}$ acceptance, similar to the one used by the COMPASS experiment at CERN.

Since the details of charm detection in experiment are not yet available,
we will perform all our calculations on the charm quark level, i.e., we do not attempt to model the hadronization of
charm quarks into charmed mesons and their subsequent decays. For the PAX experiment, 
however, identification of open charm events most
likely proceeds through the detection of decay muons for which a 
lower limit on their momentum of $p_{\mu}=1\,\mathrm{GeV}$ is
required \cite{ref:gsi-pax,ref:cont-priv}. Therefore we impose a similar cut on the
laboratory momentum of the primary charm quark in all our calculations for GSI-FAIR.

In all unpolarized calculations at LO and NLO accuracy we use the CTEQ6L1 and 
CTEQ6M parton distribution functions \cite{ref:cteq} and
the corresponding LO and NLO values for the strong coupling, respectively. 
Unless stated otherwise, the GRSV ``standard'' set
of helicity-dependent parton densities \cite{ref:grsv} is used in the computation of 
polarized cross sections and the charge and spin asymmetries.
Since $m_c=1.35\,\mathrm{GeV}$ provides a good description 
of most of the data shown in Fig.~\ref{fig:figure3}, we make this
our default choice for the charm quark mass.

\subsection{Expectations for charm production cross sections}\label{PR1}
\begin{figure}[th!]
\vspace*{-0.75cm}
\epsfig{figure=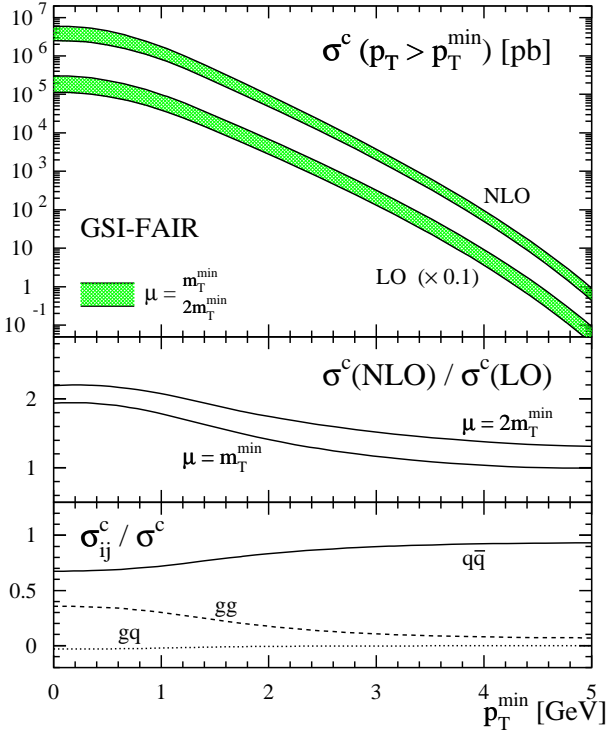,width=0.52\textwidth}
\caption{\label{fig:figure4} upper panel: LO and NLO unpolarized charm cross section at GSI-FAIR,
integrated over $p_T>p_T^{\min}$ and the angular acceptance $5^{\circ}\le \theta_c\le 130^{\circ}$,
using $m_c=1.35\,\mathrm{GeV}$.
The shaded bands indicate the uncertainties from varying $\mu=\mu_r=\mu_f$
in the range $m_T^{\min}\le\mu\le 2m_T^{\min}$;
middle panel: ratio of the NLO and LO cross sections (``K-factor'');
lower panel: fractional contribution of the different partonic channels $\sigma^c_{ij}$ to
the NLO cross section $\sigma^c$ for $\mu=\sqrt{2}m_T^{min}$.}
\end{figure}

Figure~\ref{fig:figure4} shows our expectations  for
the unpolarized charm production cross section (\ref{eq:diffxsec}) at GSI-FAIR at LO and NLO accuracy,
integrated over transverse momentum $p_T>p_T^{\min}$ and the 
angular acceptance of $5^{\circ}\le \theta_c\le 130^{\circ}$
for PAX. The shaded bands indicate the theoretical uncertainties when the factorization and renormalization
scales are varied simultaneously in the range $m_T^{\min}\le\mu_f=\mu_r\le 2m_T^{\min}$.
Also shown in Fig.~\ref{fig:figure4} are the ``K-factors'', the ratio of the NLO and LO cross sections, for two
choices of scales $\mu_f=\mu_r$ and the fractional contributions of the different partonic channels $\sigma^c_{ij}$ to
the NLO cross section.

Besides the sizable dependence on the scales $\mu_{f,r}$, there is also a 
similar uncertainty due to the choice of $m_c$ in the
region $p_T^{\min}\lesssim 1\,\mathrm{GeV}$, in line with the observations 
for the total charm yields in Fig.~\ref{fig:figure3}.
For $p_T^{\min}\gtrsim 2\,\mathrm{GeV}$, however, varying $m_c$ in the 
range $1.25\,\mathrm{GeV}\le m_c \le 1.45\,\mathrm{GeV}$ has a negligible impact
on the cross section shown in the upper panel of Fig.~\ref{fig:figure4}.
It is worth to notice that there is only a rather marginal reduction in the scale 
ambiguity when going from the LO to the
NLO approximation. This is not unexpected for experiments with limited c.m.s.\ energies, and similar observations
have been made for single-inclusive hadron production \cite{ref:jsv-fixedtarget}.
From the lower panel of Fig.~\ref{fig:figure4} one can infer that the quark-antiquark annihilation subprocess
is the most important contribution to the cross section. This can be readily understood since quarks and antiquarks
are both ``valence'' partons in the proton and antiproton, respectively, 
and from the fact that one probes fairly large momentum fractions $x_{1,2}\gtrsim 0.1$. 
The genuine NLO quark-gluon subprocess is negligible in the entire $p_T$ range shown.
%
\begin{figure}[th!]
\vspace*{-0.75cm}
\epsfig{figure=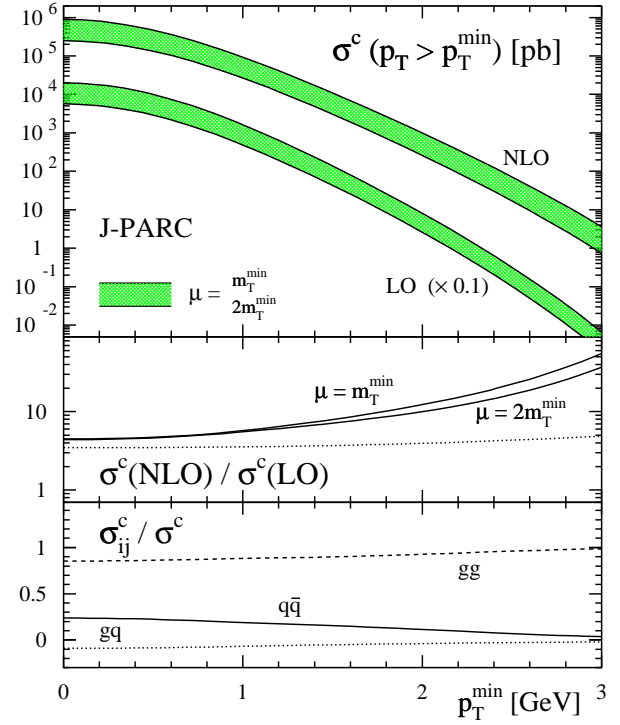,width=0.52\textwidth}
\caption{Same as in Figure \ref{fig:figure4}, but now for J-PARC kinematics.
\label{fig:figure5}}
\end{figure}

The corresponding results for J-PARC are summarized in Fig.~\ref{fig:figure5}. There are striking differences
compared to the result for GSI-FAIR shown in the previous figure. The dependence of the charm cross section on
unphysical scales $\mu_{f,r}$ is even larger here and does not improve when NLO corrections are included.
This can be taken as a strong indication that higher order terms in the perturbative series are very important.
Hopefully, a resummation of the leading terms to all orders in $\alpha_s$ will tame the scale ambiguities.
Secondly, the size of the NLO corrections compared to the LO term, displayed in the middle panel of
Fig.~\ref{fig:figure5}, seems to be beyond control. Most of the pathological behavior of the $K$-factor at
large $p_T^{\min}$ can be attributed to the differences in the LO and NLO gluon distributions 
at large $x_{1,2}$, where they are basically 
unconstrained by data \cite{ref:cteq}. If one uses NLO parton densities in the
calculation of the LO cross section, the $K$-factor does not show such a sharp rise, though it remains large
(dotted curve in Fig.~\ref{fig:figure5}).
Despite the rather large $x_{1,2}$ values probed at J-PARC, and contrary to what happens in $\bar{p}p$ collisions at
GSI-FAIR, the gluon-gluon fusion subprocess is by far the dominant mechanism to produce the charm quark.
We also note that varying $m_c$ in the range $1.25\,\mathrm{GeV}\le m_c \le 1.45\,\mathrm{GeV}$ 
has a somewhat bigger impact
on the cross sections displayed in the upper panel of Fig.~\ref{fig:figure5} than in case of GSI-FAIR due
to the smaller c.m.s.\ energy.

Figures~\ref{fig:figure4} and \ref{fig:figure5} already demonstrate 
the potential of future low energy $\bar{p}p$ and $pp$
experiments in further constraining quark and gluon distributions, 
respectively, in the medium-to-large $x$ region, difficult
to access at high energy colliders. They also show, however, that the pQCD framework can not be taken for granted in
this energy regime, and its applicability has to be carefully scrutinized first by comparing the theoretical 
expectations with data.

\begin{figure*}[th!]
\vspace*{-0.75cm}
\begin{center}
\epsfig{figure=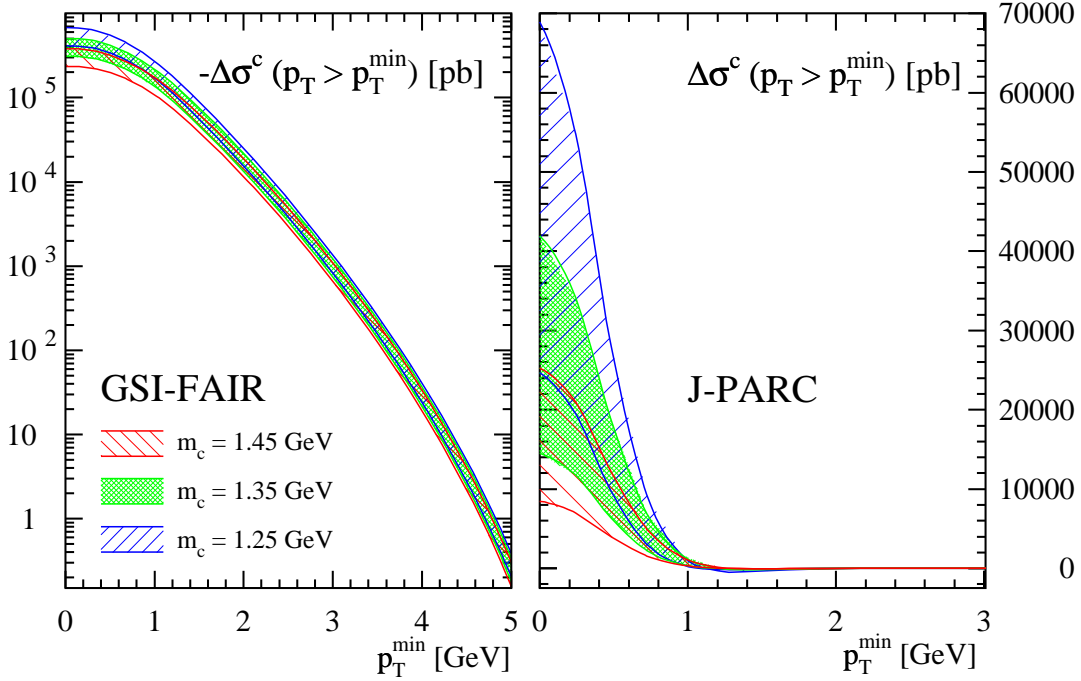,width=0.85\textwidth}
\end{center}
\vspace*{-0.5cm}
\caption{Expectations for the longitudinally polarized cross section for charm production 
at GSI-FAIR (left panel) and J-PARC (right panel), using the GRSV ``standard'' set \cite{ref:grsv}.
Note that the results on the left hand side are for $-\Delta\sigma$.
The scale uncertainty is shown for three different choices of $m_c$, varying $\mu=\mu_r=\mu_f$
in the range $m_T^{\min}\le\mu\le 2m_T^{\min}$.\label{fig:figure6}}
\end{figure*}
Figure~\ref{fig:figure6} shows the corresponding longitudinally polarized cross sections at NLO accuracy 
for GSI-FAIR and J-PARC, respectively. Instead of giving also the LO results 
[the $K$-factors are smaller than in the unpolarized case, between 1.5 and 2 (2 and 4) for GSI-FAIR (J-PARC)], 
we chose to display the dependence of the
polarized cross sections on $m_c$. Since J-PARC can cover only a smaller range in $p_T$ than GSI-FAIR due to the
smaller $\sqrt{S}$, the mass effects are more important here.
The fractional contributions of the different subprocesses are very similar to
those shown in Fig.~\ref{fig:figure4} for GSI-FAIR and strongly dependent on the size of the polarized gluon
distribution in case of J-PARC, as can be expected already from the lower panel of Fig.~\ref{fig:figure5}.
We note that at J-PARC kinematics, the polarized cross section exhibits a node at $p_T^{\min}\simeq 1\,\mathrm{GeV}$
if the GRSV ''standard'' parton densities are used in the calculation. 

%
\subsection{The unpolarized and polarized charge asymmetry \label{sec:ac}}
%
We now turn to a detailed discussion of the charge asymmetry $A_C$ defined in 
Eq.~(\ref{eq:chargeasy}) and Sec.~\ref{sec:chargeasym}. We show expectations for the size of the effect and
discuss the theoretical uncertainties due to variations of $\mu_{f,r}$ and $m_c$.
All results are presented as a function of the c.m.s.\ rapidity $y$ of the heavy (anti-)quark,
which is related to the rapidity in the laboratory frame $y_{lab}$ by a simple additive boost.
Positive rapidities refer to the direction of the anti-proton and proton beam
at GSI-FAIR and J-PARC, respectively. Recall that rapidity $y$ and pseudo-rapidity $\eta$ are not
the same for massive particles. The relation between $y$ and the scattering angle $\theta_c$ of the
heavy quark depends both on $p_T$ and $m_c$
\begin{equation}
\cos\theta_c = \left( \sqrt{\frac{m_T^2}{1-\tanh^2 y}} \tanh y\right)/\sqrt{\frac{m_T^2}{1-\tanh^2 y} - m_c^2}
\end{equation}
while $\cos \theta_c = \tanh \eta$.
Therefore, angular cuts imposed upon the heavy quarks by the experiments do not translate in 
simple, $p_T$ independent cuts for rapidity-dependent differential cross sections. 
%
\begin{figure}[bht!]
\vspace*{-0.5cm}
\epsfig{figure=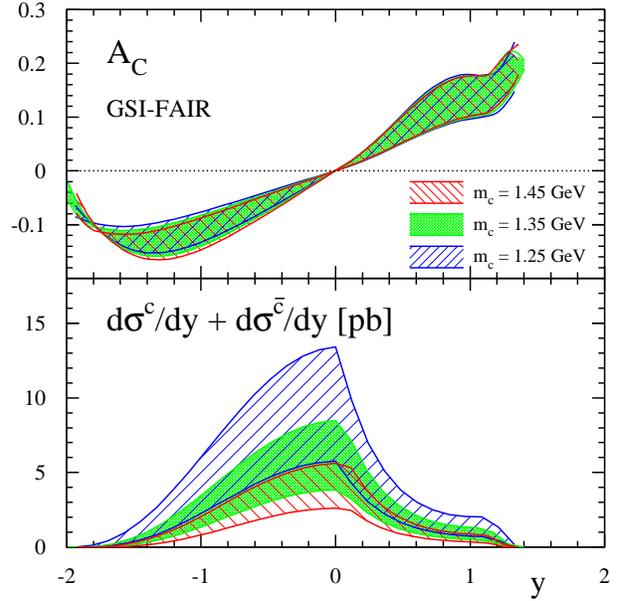,width=0.52\textwidth}
\vspace*{-0.3cm}
\caption{The unpolarized charge asymmetry $A_C$ (upper panel) and the NLO c.m.s.\
rapidity-dependent differential charm plus anti-charm cross section 
$d\sigma^c/dy +d\sigma^{\bar{c}}/dy$ (lower panel) for GSI-FAIR.
The scale uncertainty is shown for three different choices of $m_c$, varying $\mu=\mu_r=\mu_f$
in the range $m_c\le\mu\le 2m_c$. 
\label{fig:figure7}}
\end{figure}

\begin{figure}[thb!]
\vspace*{-0.5cm}
\epsfig{figure=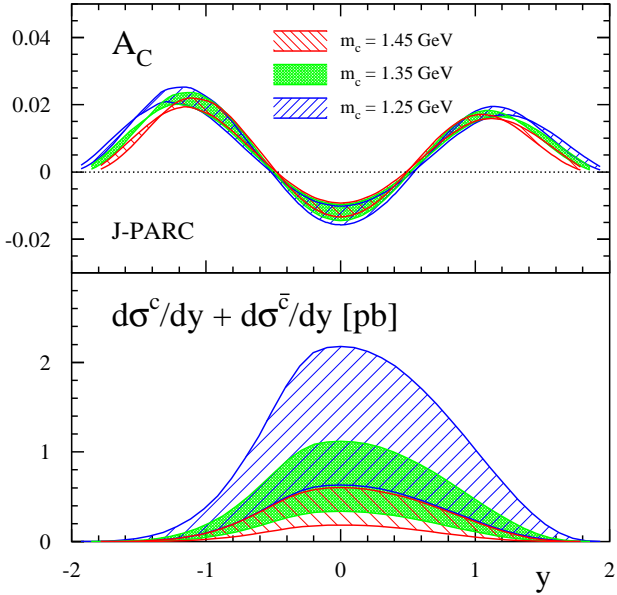,width=0.52\textwidth}
\caption{Same as in Fig.~\ref{fig:figure7} but for J-PARC kinematics.\label{fig:figure8}}
\end{figure}
The upper panels of Figs.~\ref{fig:figure7} and \ref{fig:figure8} show our 
expectations for the unpolarized charge asymmetry $A_C$ at ${\cal{O}}(\alpha_s^3)$
for charm quarks at GSI-FAIR and J-PARC, respectively, using the phenomenological inputs
and experimental acceptance cuts specified in Sec.~\ref{sec:inputs}.
The results for $A_C$ are largely independent of the choice for the charm quark mass $m_c$,
in contrast to the sizable mass dependence observed for the NLO c.m.s.\ rapidity-dependent differential
cross sections for the sum of charm and anti-charm production (lower panels), which enters
in the denominator of $A_C$ in Eq.~(\ref{eq:chargeasy}).

Note that the c.m.s.\ rapidity $y$ and the experimentally relevant rapidity $y_{lab}$
in the laboratory frame are simply related by $y_{lab}=y-0.737$ and
$y_{lab}=y+2.334$ for GSI-FAIR and J-PARC, respectively.
The scale $\mu_{f,r}$ dependence partially cancels out in $A_C$, as can be seen by comparing the
upper and lower panels of Figs.~\ref{fig:figure7} and \ref{fig:figure8}.
A residual dependence on $\mu_{f,r}$  is not surprising since 
the numerator of $A_C$ at ${\cal{O}}(\alpha_s^3)$ is effectively a
LO approximation. It vanishes at ${\cal{O}}(\alpha_s^2)$ and is
free of collinear singularities. 
Also note that the sharp drop of $d\sigma^c/dy +d\sigma^{\bar{c}}/dy$ in the lower
panel of Fig.~\ref{fig:figure7} is due to the cut imposed on the momentum of the 
heavy (anti-)quark, $p_{lab}>1\,\mathrm{GeV}$.
%
\begin{figure}[thb!]
\vspace*{-0.5cm}
\epsfig{figure=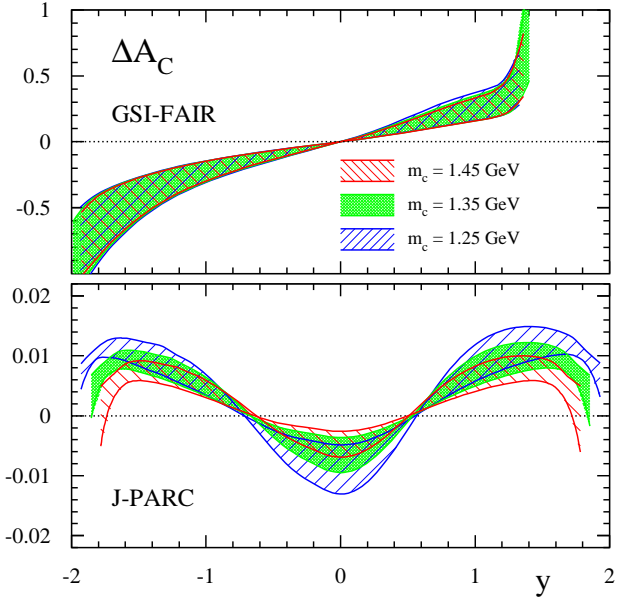,width=0.52\textwidth}
\caption{Same as in the upper panels of Figs.~\ref{fig:figure7} and \ref{fig:figure8} but
now for the polarized charge asymmetry $\Delta A_C$. \label{fig:figure9}}
\end{figure}
\begin{figure}[hbt!]
\vspace*{-0.5cm}
\epsfig{figure=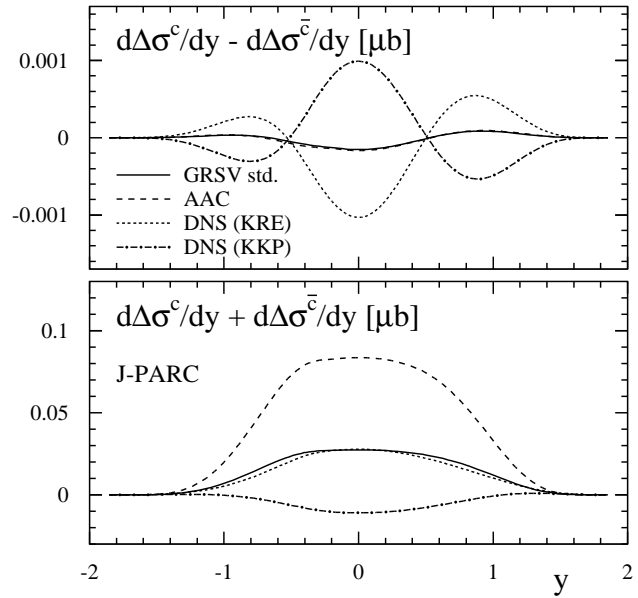,width=0.52\textwidth}
\caption{Numerator (upper panel) and denominator (lower panel) 
of the polarized charge asymmetry $\Delta A_C$ at J-PARC, calculated with 
different sets of polarized parton densities. \label{fig:figure10}}
\end{figure}

Figure~\ref{fig:figure9} shows our results for the longitudinally polarized charge asymmetry $\Delta A_C$ defined
as in Eq.~(\ref{eq:chargeasy}), but with all cross sections $d\sigma$ replaced by their helicity dependent
counterparts $d\Delta \sigma$. Again, the scale uncertainty is shown for three different 
choices of $m_c$, varying $\mu=\mu_r=\mu_f$ in the range $m_c\le\mu\le 2m_c$.
All results are obtained with the GRSV ``standard'' set \cite{ref:grsv} of spin-dependent parton densities.
As in the unpolarized case, the mass dependence largely drops out in $\Delta A_C$,
in particular for GSI-FAIR,  but a residual scale $\mu_{f,r}$ ambiguity remains.

\begin{figure*}[th!]
\vspace*{-0.75cm}
\begin{center}
\epsfig{figure=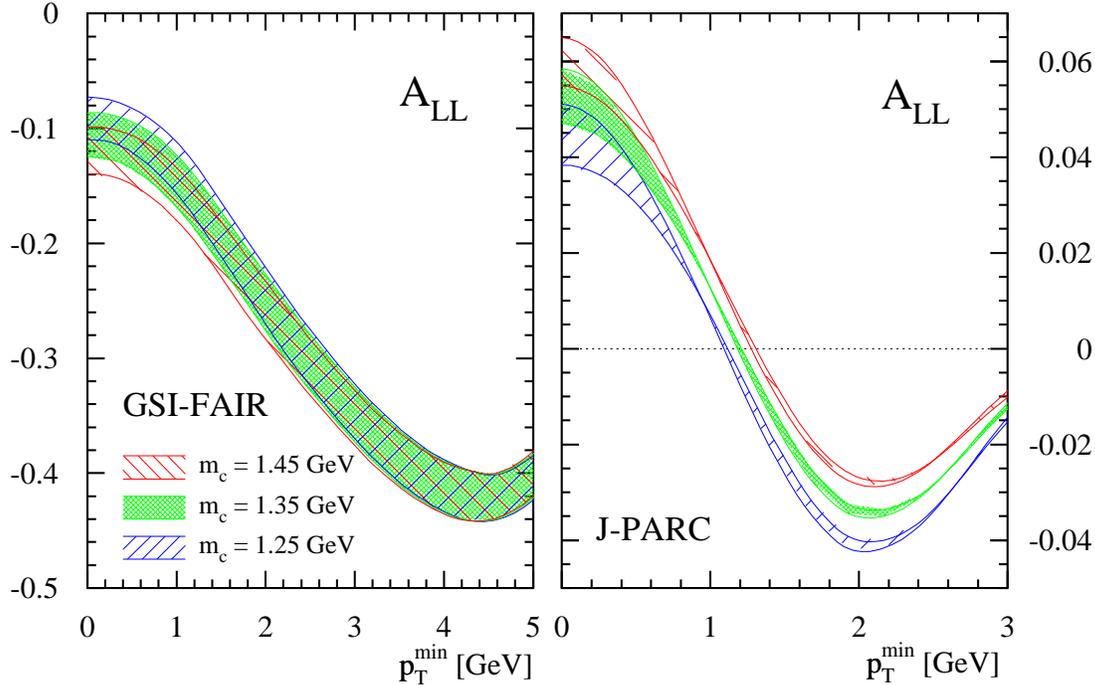,width=0.85\textwidth}
\end{center}
\vspace*{-0.5cm}
\caption{The integrated double-spin asymmetry $A_{LL}$ as function of $p_T^{\min}$ 
at NLO accuracy for GSI-FAIR (left panel) and J-PARC (right panel) using the GRSV ``standard''
distributions and the same experimental cuts as before. 
The scale uncertainty is shown for three different choices of $m_c$, varying $\mu=\mu_r=\mu_f$
in the range $m_T^{\min}\le\mu\le 2m_T^{\min}$. \label{fig:figure11}}
\end{figure*}
Since the mass and scale dependence of $d\Delta\sigma^c/dy +d\Delta\sigma^{\bar{c}}/dy$ is qualitatively
very similar to the corresponding unpolarized cases shown in the lower panels of Figs.~\ref{fig:figure7}
and \ref{fig:figure8}, we refrain from giving these results here.
Instead, we shall discuss the significant dependence of both numerator and denominator of $\Delta A_c$ on the
choice of a particular set of polarized parton densities. 

Figure~\ref{fig:figure10} shows the 
numerator and the denominator of the polarized charge asymmetry $\Delta A_C$ at J-PARC, obtained with 
different sets of polarized parton densities. Apart from our default set, GRSV ``standard'', 
we also use the sets of AAC \cite{ref:aac} and DNS \cite{ref:dns}. The latter is based on an
analysis using also data from polarized semi-inclusive deep-inelastic scattering. Two different choices
of parton-to-hadron fragmentation functions have been made in the DNS analysis, and the two resulting
sets, labeled as DNS (KRE) and DNS (KKP) in Fig.~\ref{fig:figure10}, differ mainly in the
sea quark content, in particular, $\Delta\bar{u}$, which has opposite sign in both sets. 
The positive polarization of $\Delta\bar{u}$ in DNS (KKP), unlike in all others sets
of spin-dependent parton densities, is responsible for the opposite sign
of $d\Delta\sigma^c/dy \pm d\Delta\sigma^{\bar{c}}/dy$ obtained with DNS (KKP).
In the denominator of $\Delta A_C$, gluon-gluon fusion does not 
drop out and can make a significant contribution depending
on the amount of gluon polarization $\Delta g$. The set with the largest $\Delta g$, AAC, gives the largest
cross section. The other three sets have relatively small gluon distributions and quark-antiquark annihilation
is equally important.
On the other hand, different sets of polarized parton densities have only very limited impact on the results for
$d\Delta\sigma^c/dy \pm d\Delta\sigma^{\bar{c}}/dy$ and $\Delta A_C$ for GSI-FAIR. This can be expected, since
in $\bar{p}p$ collisions at small $\sqrt{S}$ one predominantly probes 
the fairly well constrained valence quark distributions,
and uncertainties in the polarized sea-quark and gluon densities do not matter much.

Finally, we note that {\em without} taking into account any experimental cuts, 
$A_C$ and $\Delta A_C$ are anti-symmetric functions
in the c.m.s.\ rapidity for $\bar{p}p$ collisions, as can be anticipated from 
Figs.~\ref{fig:figure7} - \ref{fig:figure9} 
(which, however, do include certain acceptance cuts).
Similarly, $A_C$ and $\Delta A_C$ are symmetric in c.m.s.\  rapidity for $pp$ experiments.
This is a consequence of the anti-symmetric and symmetric initial states $\bar{p}p$ and $pp$, respectively.
The charge asymmetry on the partonic level for the dominant $q\bar{q}$ subprocess 
implies that $Q$ is preferentially emitted into the direction of $q$ and $\bar{Q}$ into the direction of $\bar{q}$.
As was explained in \cite{ref:kuhn}, in $pp$ collisions one then finds an excess of centrally produced $\bar{Q}$, while
$Q$ dominates at large absolute rapidities. This is also what we observe for J-PARC in Figs.~\ref{fig:figure8} and
\ref{fig:figure9}.
The size of $A_C$ and $\Delta A_C$ for $pp$ collisions at J-PARC is significantly smaller than 
for $\bar{p}p$ collisions at GSI-FAIR, simply because of the fact that for the relevant $q\bar{q}$ subprocess
both partons are valence quarks in $\bar{p}p$, greatly enhancing its relative contribution.
When integrated over rapidity and without any kinematical restrictions, $A_C$ and $\Delta A_C$ vanish, 
and the total yields of charm and anti-charm quarks are the same.
\begin{figure*}[th!]
\vspace*{-0.75cm}
\begin{center}
\epsfig{figure=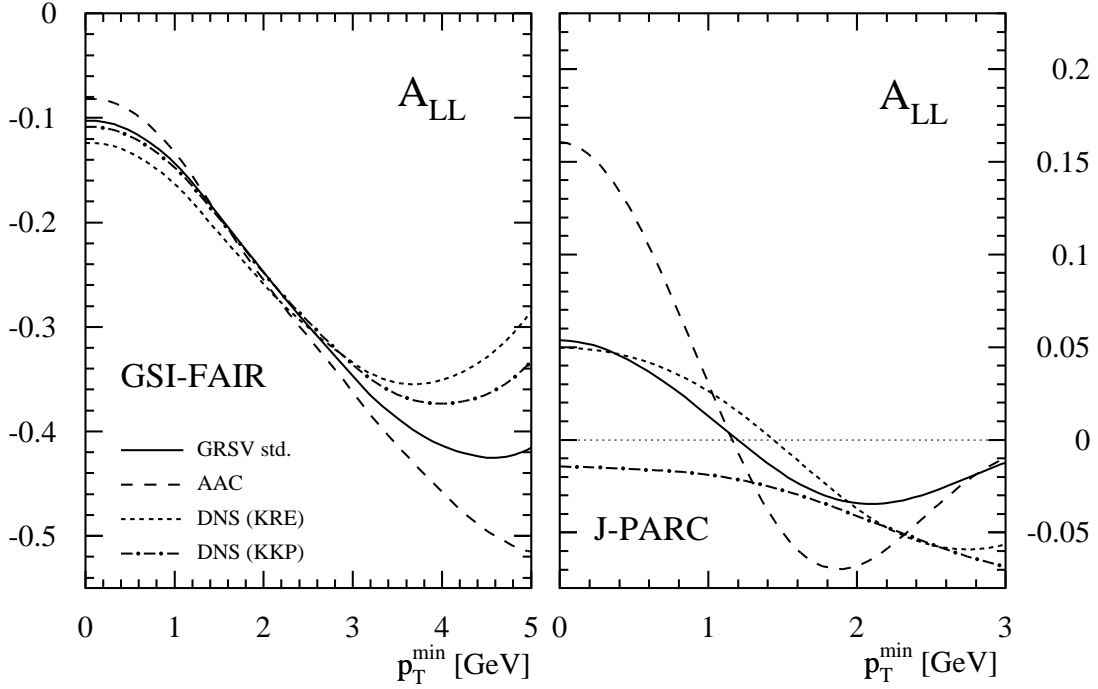,width=0.85\textwidth}
\end{center}
\vspace*{-0.5cm}
\caption{As in Fig.~\ref{fig:figure11} but now comparing the results obtained with different
sets of polarized parton densities using $m_c=1.35\,\mathrm{GeV}$ and $\mu_f=\mu_r=\sqrt{2}m_T^{\min}$.
\label{fig:figure12}}
\end{figure*}
%

\subsection{Expectation for longitudinal spin asymmetries \label{sec:all}}
%
\begin{figure}[thb!]
\vspace*{-0.5cm}
\epsfig{figure=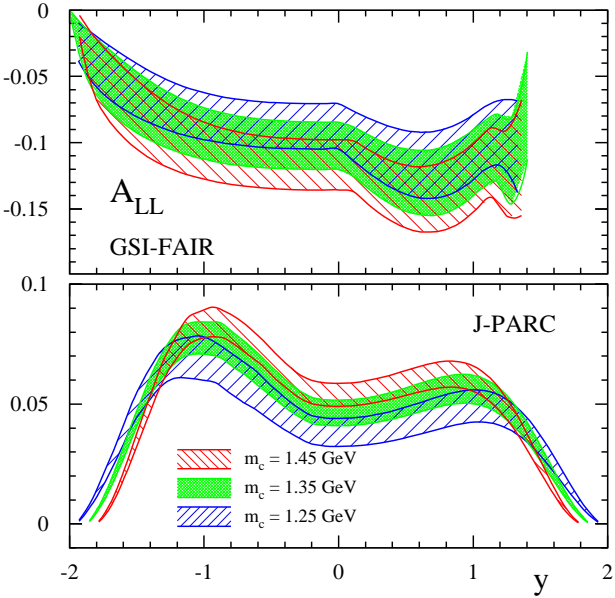,width=0.52\textwidth}
\caption{As in Fig.~\ref{fig:figure11} but now as function of the c.m.s.~rapidity $y$.
The scales $\mu_{f,r}$ are varied in the range $m_c\le \mu_f=\mu_r\le 2m_c$.
\label{fig:figure13}}
\end{figure}

Experiments usually present their spin-dependent measurements in terms of spin-asymmetries rather than
polarized cross sections. The double-spin asymmetry $A_{LL}$, defined in Eq.~(\ref{eq:all}),
has the advantage that many experimental uncertainties cancel in the ratio, in particular, it is not 
required to determine the absolute normalization of the helicity dependent cross sections.
Also, theoretical uncertainties may cancel to some extent in $A_{LL}$. However, before exploiting this, one has
to make sure that pQCD is applicable in the relevant kinematical regime by comparing, for instance, the 
unpolarized cross section with data.

Figure~\ref{fig:figure11} shows the integrated double-spin asymmetry $A_{LL}$ as function of $p_T^{\min}$ 
at NLO accuracy for GSI-FAIR (left panel) and J-PARC (right panel), using the GRSV ``standard''
distributions and the same experimental cuts as for the underlying cross sections shown 
in Figs.~\ref{fig:figure4} - \ref{fig:figure6}.
As usual, the scale uncertainty is shown for three different choices of $m_c$, varying $\mu=\mu_r=\mu_f$
in the range $m_T^{\min}\le\mu\le 2m_T^{\min}$. As can be seen, there is still a significant scale 
ambiguity and, in case of J-PARC, also a dependence on $m_c$.
We refrain from showing LO results, but we note that because NLO corrections tend to be larger in
the unpolarized case, as discussed above, $A_{LL}$ is typically reduced by a factor of about two when
NLO corrections are included.

The sensitivity of $A_{LL}$ to different sets of polarized parton distributions is studied in
Fig.~\ref{fig:figure12}. As expected, the differences are small for GSI-FAIR which mainly
probes the fairly well known valence distributions. Only at large $p_T^{\min}$, which corresponds to
currently unexplored momentum fractions $x_{1,2}\to 1$, some differences are noticeable. 
At J-PARC, expectations for $A_{LL}$ depend much more on the choice of spin-dependent
parton densities, in line with the observations already made in the lower panel of Fig.~\ref{fig:figure10}.

Finally, in Fig.~\ref{fig:figure13} we present the spin asymmetry as a function of the c.m.s.\ rapidity $y$ and
integrated over transverse momentum. The behavior of $A_{LL}$ for GSI-FAIR in the upper panel for positive $y$
is driven by the cut on the charm momentum $p_{lab}>1\,\mathrm{GeV}$. Both mass and scale uncertainties
do not cancel and remain significant.

\section{Conclusions}
%
We have performed a detailed study of the physics opportunities with open charm production at
low c.m.s.\ energy $\bar{p}p$ and $pp$ collisions at GSI-FAIR and J-PARC, respectively, including
unpolarized and polarized cross sections and charge and spin asymmetries.
All calculations are done at ${\cal{O}}(\alpha_s^3)$ accuracy, and theoretical uncertainties due
to the choice of scales $\mu_{f,r}$ and the charm mass $m_c$ are discussed in detail.
In general, they turn out to be significant with the exception of the mass dependence
of the charge asymmetries $A_C$ and $\Delta A_C$.

Measurements of all these quantities would further our understanding of the perturbative
QCD framework and the nucleon structure expressed in terms of unpolarized and polarized parton densities.
The latter are probed at large momentum fractions $x_{1,2}$, which are difficult to access at
high-energy colliders. The charge asymmetry vanishes at ${\cal{O}}(\alpha_s^2)$ and hence is
a clear probe of non-trivial QCD dynamics beyond the LO. Detailed comparisons between our
theoretical expectations for unpolarized charm yields and future data will reveal to what
extent perturbative methods are applicable in hadron-hadron collisions at $\sqrt{S}\approx 10\div 15\,\mathrm{GeV}$
and will open up a window to perturbative resummations and/or the transition into the
non-perturbative regime so far only little explored.

\section*{Acknowledgements}
We are grateful to Marco Contalbrigo providing 
information about the GSI-FAIR PAX project.
J.R.\ was supported by a grant of the ``Cusanuswerk'', Bonn, Germany.
This work was supported in part by the BMBF, Germany.


%
\end{document}